\documentclass[reprint,amsmath,amssymb,aps,floatfix]{revtex4-1}

\usepackage{epstopdf}
\usepackage{graphicx}
\usepackage{siunitx}
\usepackage{amsmath}
\usepackage{makecell}

\begin{document}
\title{Experimental and theoretical characterisation of Stokes polarimetry of the potassium D1 line with neon buffer gas broadening}

\author{Sharaa A. Alqarni, Danielle Pizzey, Steven Wrathmall and Ifan G Hughes}

\affiliation{Department of Physics, Durham University, South Road, Durham, DH1 3LE, UK}

\affiliation{Corresponding author: danielle.boddy@durham.ac.uk}

\begin{abstract}
This study presents a comprehensive experimental and theoretical characterisation of Stokes polarimetry in potassium (K) vapour on the D1 line. Measurements were performed in the weak-probe regime, investigating the influence of neon buffer gas in the presence of an applied magnetic field in the Faraday geometry. While previous Stokes polarimetry studies in alkali-metal vapours have been conducted, the specific effects of buffer gas-induced broadening and shifts on the observed Stokes parameters remained largely underexplored. Here, experimental measurements of absolute absorption and dispersion were compared with a theoretical model for the electric susceptibility of the vapour, calculated using the established software package $ElecSus$. This work marks the first application of $ElecSus$ to model buffer gas polarimetry of the potassium D1 line, with validation performed against experimental spectra for magnetic fields up to 1.2~kG. Our findings provide new insight into how the presence of buffer gas influences the observed Stokes parameters, thereby enhancing the predictive capabilities of theoretical frameworks for atom-light interactions in buffer-gas environments.
\end{abstract}

%
%
%
\maketitle
%
%

\section{Introduction}

The interaction of light with atomic systems underpins diverse fields, from quantum optics and metrology \cite{PezzeRMP} to astrophysics \cite{erdelyi2022solar,vargas2022mesosphere,solar4} and medical imaging \cite{zhang2020recording, boto2017new}. A comprehensive understanding of this interaction, provides invaluable insight into an atom's fundamental properties, internal dynamics, and response to external perturbations \cite{cohen}. with the light's polarisation carrying information about the atomic ensemble's quantum state.

Stokes polarimetry, which measures the four Stokes parameters ($S_{0}, S_{1}, S_{2}, S_{3}$ or $I, Q, U, V$), fully describes an electromagnetic wave's polarisation state. When applied to light that has interacted with an atomic vapour, Stokes polarimetry serves as a powerful diagnostic tool \cite{vectorlight}. It reveals subtle changes in the atomic medium induced by factors such as applied magnetic fields (e.g. though the Zeeman or Hanle effects), electric fields, or collisional processes \cite{LandiDeglInnocenti}. These polarisation signatures are highly sensitive to the atomic environment, making polarimetry an exquisite probe for fundamental atomic physics studies~\cite{budker2002resonant, Budker2007:magnetometry}.

There is much current interest in using thermal atomic vapours as media for both fundamental physics and applications; see, for example, the review articles in the Special Issue \textit{Focus in Hot Atomic Vapors} in New Journal of Physics~\cite{fabricant2023build,finkelstein2023practical,downes2023practical,glorieux2023hot,alaeian2024manipulating,zhang2024interplay,mausezahl2024tutorial, uhland2023build, pizzey2022laser}. Alkali-metal atoms, with their relatively simple electronic structure and well-separated D-line transitions, serve as excellent testbeds for both theoretical models and experimental investigations of atom-light interactions. For instance, Stokes polarimetry studies have successfully probed complex atom-light interactions in rubidium vapour, even in the presence of large magnetic fields and high densities \cite{Weller_2012_stokes, ponciano2020absorption}. However, while these investigations advanced our understanding of light propagation in such media, they did not specifically explore how buffer gas presence influences the observed Stokes parameters through broadening and shifts. Our work focuses on potassium (K), which offers strong D1 (4$^{2}$S$_{1/2}$~$\rightarrow$~4$^{2}$P$_{1/2}$ at 769.9~nm) and D2 (4$^{2}$S$_{1/2}$~$\rightarrow$~4$^{2}$P$_{3/2}$ at 766.5~nm) transitions. These lines are readily accessible with diode lasers and are sensitive to external magnetic fields and collisional effects.
Another motivation for the study of potassium is
the existence of both fermionic and bosonic isotopes;
this allows the study of, for example, interactions in Bose–Fermi mixtures~\cite{sowinski2019one, yan2024collective} or testing of different sub-Doppler cooling mechanisms~\cite{sievers2015simultaneous, fernandes2012sub, nath2013quantum}.

 The utility of spectroscopic study of thermal atomic vapours extends to practical applications; for example, buffer-gas filled potassium vapour cells are employed in solar observations to image the Sun through the atomic medium from which the four Stokes parameters can be measured, enabling the extraction of solar magnetograms \cite{erdelyi2022solar}. However, despite this practical relevance, these prior investigations---whether in rubidium or in solar application---did not specifically explore how buffer gas presence influences the observed Stokes parameters through broadening and shifts. 

In vapour-cell experiments, buffer gases are often included to control atomic diffusion, extend interactions times \cite{brandt1997buffer,vanier2005atomic, Kroemer:15}, or investigate the impact of collisions on atomic properties \cite{allard1982effect,PhysRevLett.108.183202}. Indeed, nominally ``buffer-gas free'' vapour cells can contain sufficient gas to modify the optical properties of the medium~\cite{Higgins_2024}. These inert gas collisions can induce significant effects such as pressure broadening, spectral shifts, and, crucially for polarimetric studies, depolarisation of atomic states \cite{rotondaro1997collisional,rotondaro1998collision,pitz2012pressure,pitz2009pressure,zameroski2011pressure,andalkar2002high}. Although the influence of common buffer gases such as helium and argon on alkali atom-light interactions is well-studied~\cite{corney1978atomic, lewis1980collisional, thorne1999spectrophysics}, detailed experimental characterisation of neon as a buffer gas with potassium is less comprehensively documented \cite{pitz2012pressure,li2016pressure, Alqarni_2025}. Such studies are vital for a complete understanding of collisional effects across different noble gas species and for providing valuable data for theoretical models. 

To complement experimental measurements, computational models are essential for quantitatively understanding and predicting atom-light interactions. For alkali-metal atoms, tools such as $ElecSus$ provide a robust framework for simulating the transmission and polarisation of light through atomic vapours \cite{Zentile2015b, keaveney2018elecsus}. It accurately incorporates effects such as Doppler broadening, natural linewidth, hyperfine structure, and the presence of external magnetic fields. By offering detailed theoretical predictions of Stokes profiles under various conditions, $ElecSus$ enables direct comparison with experimental data, facilitating the extraction of experimental parameters and validating theoretical models.

This paper presents an experimental investigation of Stokes polarimetry of potassium atomic vapour in the presence of neon buffer gas. Our measurements characterise the full polarisation state of resonant D1 laser light transmitted through the atomic medium. These experimental results are compared with theoretical predictions generated using the computational model $ElecSus$. Our aim is to provide new insight into how neon buffer gas influences the observed Stokes parameters, thereby contributing to the fundamental understanding of atom-light interactions in buffer-gas environments and enhancing the predictive capabilities of theoretical frameworks such as $ElecSus$.

We perform the experiment in the Hyperfine Paschen Back (HPB) regime, which is achieved when the Zeeman shift from the external magnetic field exceeds the ground state hyperfine interaction. 
An estimate of the field strength required to reach this regime is $B_{\rm HPB}=A_{\rm hf}/\mu_{\rm B}$, where $\mu_{\rm B}$ is the Bohr magneton, and $A_{\rm hf}$ the magnetic dipole constant for the ground term; this is evaluated as
165~G~\cite{sargsyan2018selective, Alqarni_2025} for $^{39}$K. In this study the applied laboratory fields are in the range 750--1200 G; as these far exceed $B_{\rm HPB}$ the spectroscopy is performed deep into the HPB regime where the results are easier to interpret. Magneto-optical filters and wing selectors for solar filter experiments are also performed with similar (or larger) field stengths~\cite{hale2020modelling, refId1,refId2, giovannelli2020tor, erdelyi2022solar}. 
 
Recently we investigated the effect of buffer gas and magnetic field for varying amounts of neon as a buffer gas in potassium on the D1 transition exploring the Doppler and collisional effects on the spectrum~\cite{Alqarni_2025}. The extinction of the light traversing the atomic medium is governed by the imaginary component of the electric susceptibility. The real part of the susceptibility describes the retardation of the light passing through the medium. In the presence of an axial magnetic field (the so-called Faraday geometry) the two eigenmodes of polarisation are left and right-hand circular polarisation~\cite{briscoe2024indirect}. The presence of the field renders the medium birefringent and dichroic. As a consequence, the polarisation state of the light evolves as it traverses the medium. Stokes polarimetry is an ideal technique for characterising the polarisation state of the light exiting the cell. Both real and imaginary components need to be taken into account to describe the evolution in polarisation. For many applications including fabricating devices--especially magneto-optical filters--both the real and imaginary part of the susceptibility have to be incorporated into the theoretical treatment; the motivation for this work is to build on the characterisation of the imaginary component in~\cite{Alqarni_2025} and to study experimentally both components as a function of applied magnetic field and temperature in a regime where collisional broadening can not be neglected.

The remainder of this paper is structured as follows. We begin in Section~\ref{sec:ther_code} by outlining the theoretical framework used to analyse the temperature dependence of the four Stokes parameters of potassium vapour in the presence of a buffer gas. Section \ref{sec:exp_details} describes the experimental set-up and procedures employed for performing potassium Stokes polarimetry as a function of temperature and magnetic field strength. In Section~\ref{sec:exp_results}, we present and discuss the experimental results obtained for the potassium D1 line, comparing them with our theoretical predictions. Finally, Section~\ref{sec:conclusion} offers our conclusions and an outlook for future work.

\section{Theoretical model}
\label{sec:ther_code}

\begin{figure*}
  \centering
    \includegraphics[width=0.8\linewidth]{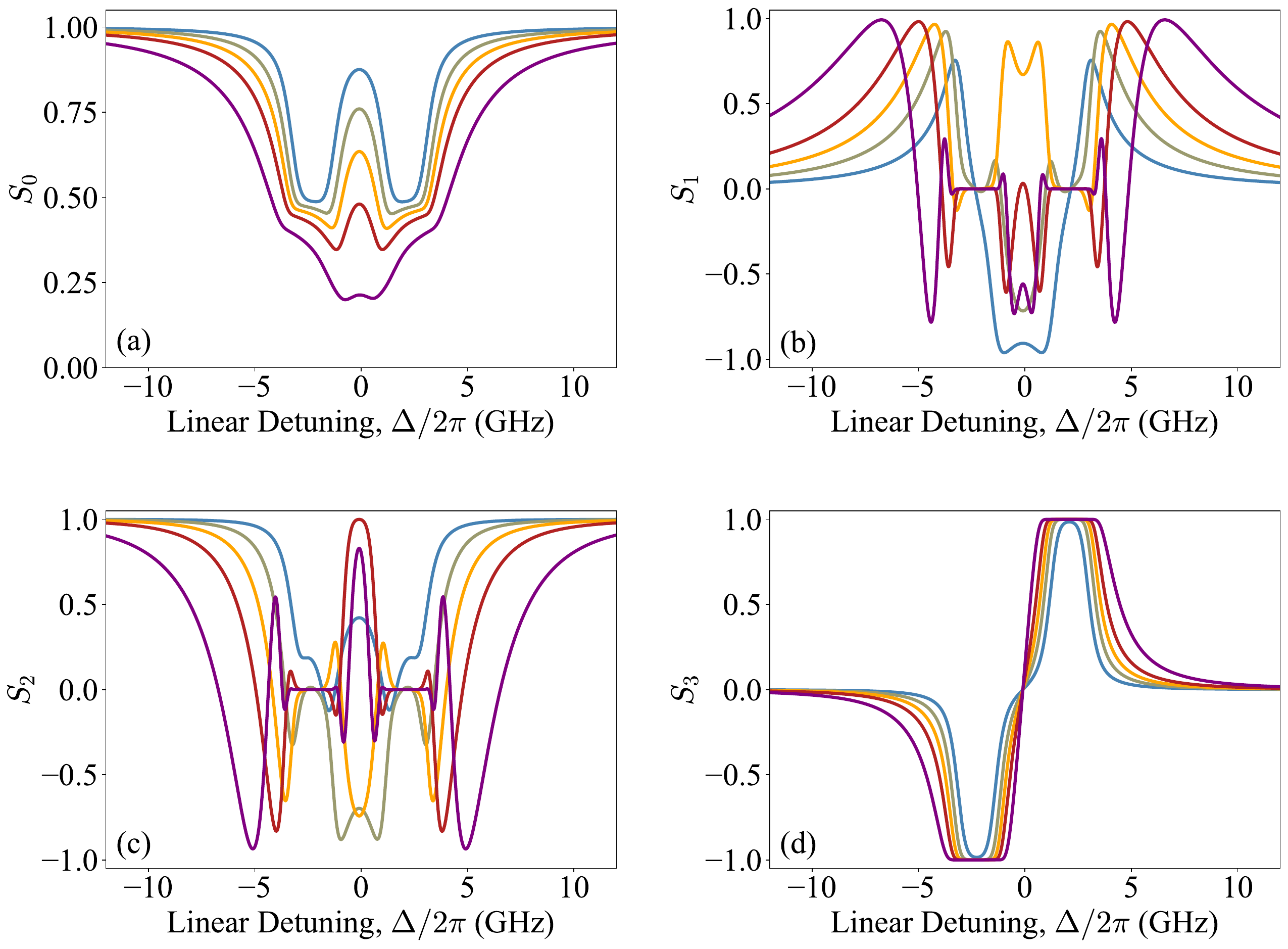}
    \caption{Theoretical $ElecSus$ calculations of the normalised Stokes parameters -- (a) $S_{0}$, (b) $S_{1}$, (c) $S_{2}$ and (d) $S_{3}$ -- for a potassium vapour cell of length 25 mm containing 60 Torr of neon buffer gas. Before the cell the polarisation state of the light is linear diagonally polarised. The applied magnetic field is 1160 G, with the vapour cell stem temperature, $T_{\rm{s}}$, increasing from 75$^{\circ}$C (blue) to 135$^{\circ}$C (purple), in 15$^{\circ}$C increments. A linear detuning of 0 GHz is the potassium line centre in the pressence of no buffer gas~\cite{hanley2015absolute}.}
    \label{fig:theoryKStokes}
\end{figure*}

\begin{figure*}
    \centering
    \includegraphics[width=0.8\linewidth]{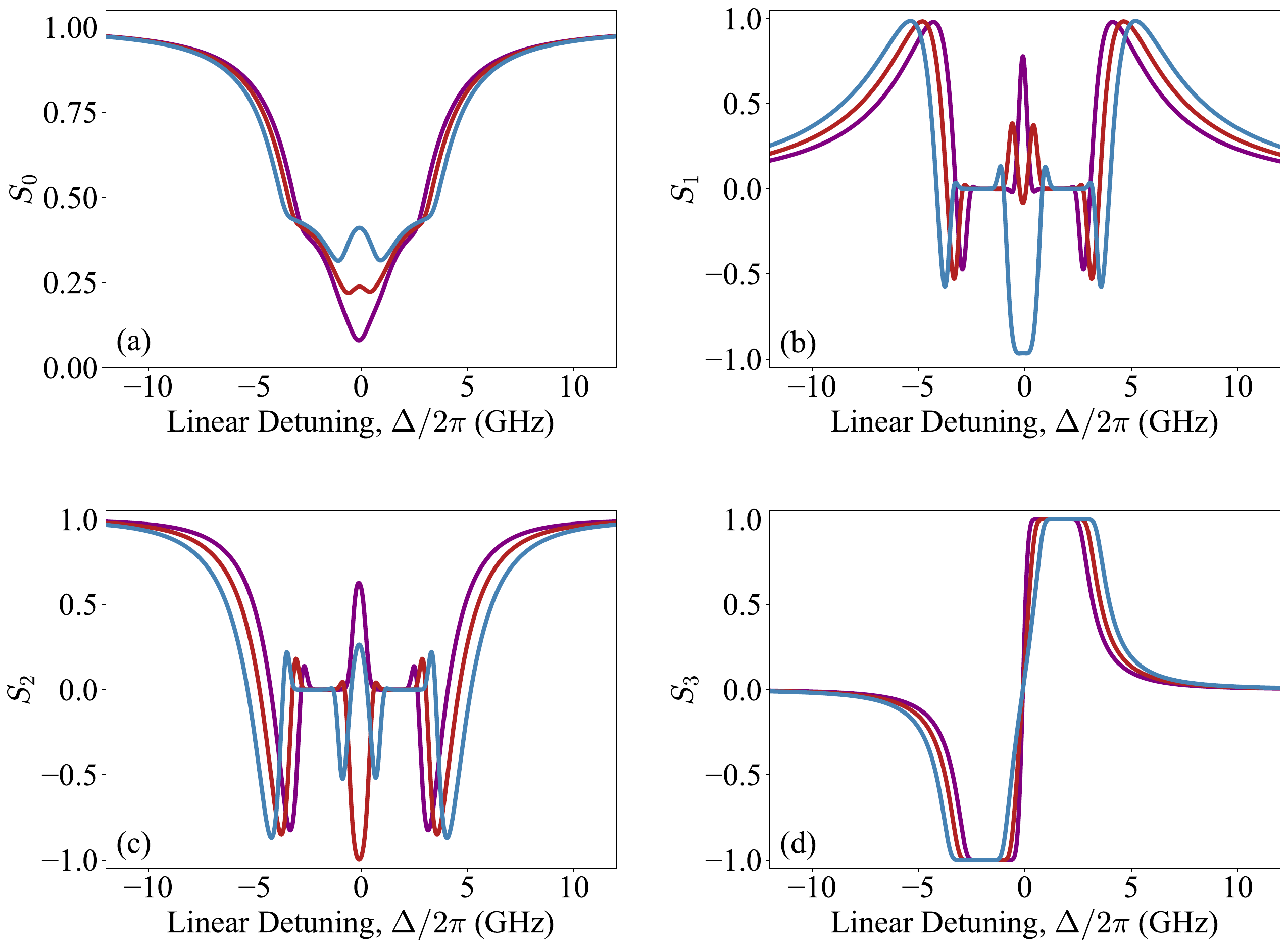}
    \caption{Theoretical $ElecSus$ calculations of the normalised Stokes parameters -- (a) $S_{0}$, (b) $S_{1}$, (c) $S_{2}$ and (d) $S_{3}$ -- for a potassium vapour cell of length 25 mm containing 60 Torr of neon buffer gas. Before the cell the polarisation state of the light is linear diagonally polarised. The vapour cell stem temperature, $T_{\rm{s}}$, is set to 120$^{\circ}$C and the magnetic field is 750~G (purple), 1000~G (red) and 1200~G (blue).}
    \label{fig:theoryKStokes_magnetic}
\end{figure*}

\begin{figure*}
    \centering
    \includegraphics[width=0.9\linewidth]{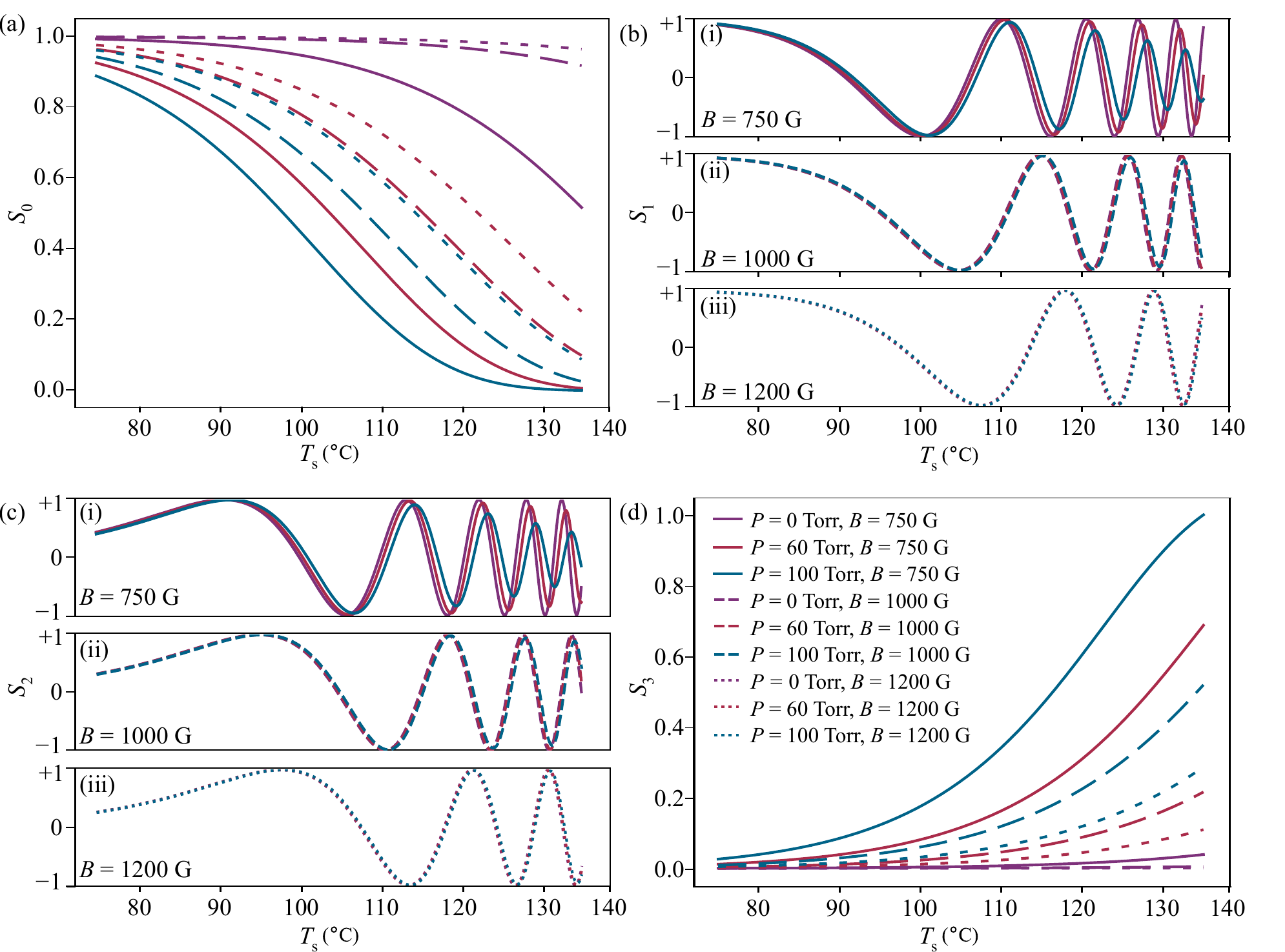}
    \caption{Theoretical $ElecSus$ calculations of the on-resonance (i.e. $\Delta/2\pi$~=~0~GHz) normalised Stokes parameter magnitude for (a) $S_{0}$, (b) $S_{1}$, (c) $S_{2}$ and (d) $S_{3}$ for a vapour cell stem temperature range between 75--135$^{\circ}$C and magnetic fields, in Faraday geometry, of 750~G (solid lines), 1000~G (dashed lines), and 1200~G (dotted lines) for three neon buffer gas filling pressures of 0~Torr (purple), 60~Torr (red), and 100~Torr (blue). }
    \label{fig:Onresdep}
\end{figure*}

This section outlines the theoretical framework employed to model the interaction of light with potassium vapour in the presence of a buffer gas, primarily drawing upon the $ElecSus$ software package~\cite{Zentile2015b,keaveney2018elecsus} with buffer-gas collisions included~\cite{Alqarni_2025}. We operate in the weak-probe regime, where the laser intensity is sufficiently low to avoid saturation effects, ensuring that the light field does not significantly alter the atomic ground-state population~\cite{doi:10.1119/1.1652039,weakprobe}. Under this assumption, the propagation of light through the atomic medium is governed by the complex electric susceptibility, $\chi\,(\omega)$, which is calculated by $ElecSus$ based on the atomic structure and environmental conditions. 

The $ElecSus$ model for alkali atoms accurately accounts for crucial physical effects, such as the hyperfine structure, Zeeman effect, Doppler broadening, and collisional broadening and shift. From the calculated complex electric susceptibility $\chi\,(\omega)$, which is a tensor quantity for polarised light in a magnetic field, $ElecSus$ determines the refractive index and absorption coefficient for different polarisation components \cite{jacknonorth}. The propagation of light through the vapour cell is modelled using a transfer matrix approach, from which the Stokes parameters of the transmitted light can be derived. 

The complete characterisation of the polarisation state of light necessitates three independent parameters: two amplitudes corresponding to orthogonal directions, and a relative phase between them~\cite{born2013principles}. These three parameters, which define the electric field, can be derived from three pairs of light intensity measurements. The polarisation state of the transmitted light is conventionally and fully characterised by the four Stokes parameters~\cite{stokes1851composition,landolfi2004polarization, schaefer2007measuring, auzinsh2010optically}. The widespread adoption of the Stokes parameters is attributed to their direct correlation with practical light-polarisation intensity measurements. These measurements are performed by recording the intensity of light along orthogonal axes across three distinct bases: (i) horizontal and vertical linear polarisation ($I_{\rm{H}}$ and $I_{\rm{V}}$, respectively); (ii) diagonal and anti-diagonal linear polarisation ($I_{\nearrow}$ and $I_{\searrow}$, respectively); and (iii) right and left-circular polarisation ($I_{\text{RCP}}$ and $I_{\text{LCP}}$, respectively).

The intensity of the transmitted light is independent of the measurement basis:
\begin{equation}
I_{\rm{T}} \equiv I_{\rm{H}} + I_{\rm{V}} = I_{\nearrow} + I_{\searrow} = I_{\text{RCP}} + I_{\text{LCP}}~.
\label{eq:StokesParameter}
\end{equation}
 
We define the normalised Stokes parameter $S_{0}$ as the ratio of the light transmitted by the medium to the incident light, $S_{0}=I_{\rm{T}}/I_0$. Measuring the light transmitted by the medium involves placing a detector after the vapour cell. This signal inherently incorporates intensity losses due to the surfaces of the vapour cell and any other optics in the path before the detector. Conversely, measuring the incident intensity, $I_{0}$, with a detector placed before the vapour cell does not account for these subsequent losses. A more accurate determination of the incident intensity involves tuning the laser significantly off-resonance and defining the raw detector signal at this frequency as $I_{0}$. Alternatively, one can scan over a broad frequency range centred on the atomic resonance frequency and take the maximum repeatable off-resonance detector signal as the $I_{0}$ value.
 
The remaining Stokes parameters, normalised with respect to the intensity of the transmitted light, $S_{1}$, $S_{2}$ and $S_{3}$, are defined as,

\begin{equation}
S_1 \equiv (I_{\rm{H}} - I_{\rm{V}}) \,/\, (I_{\rm{H}} + I_{\rm{V}}),
\end{equation}
\begin{equation}
S_2 \equiv (I_{\nearrow} - I_{\searrow}) \,/\,(I_{\nearrow} + I_{\searrow}),
\end{equation}
\begin{equation}
S_3 \equiv (I_{\text{RCP}} - I_{\text{LCP}})\,/\,(I_{\text{RCP}} + I_{\text{LCP}}).
\label{eq:Stokesparameters}
\end{equation}

Figure~\ref{fig:theoryKStokes} presents the theoretically predicted Stokes parameters ($S_{0}, S_{1}, S_{2}, S_{3}$) as a function of linear detuning (where the weighted line-centre is taken as $\Delta/2\pi = 0~\mathrm{GHz}$). The vapour cell is 25~mm long, contains potassium vapour and 60 Torr of neon buffer gas, and is subject to a magnetic field of 1200 G in a Faraday configuration; the maximum permissible field in our investigation \cite{Alqarni_2025}. The calculations were performed using $ElecSus$ in the weak-probe regime and the temperature is increased from 75$^{\circ}$C to 135$^{\circ}$C in 15$^{\circ}$C increments. The incident light is linearly polarised at 45$^{\circ}$ to the axis of polarisation, meaning there are equal measures of left- and right-circularly polarised light incident on the atomic medium. 

Figure~\ref{fig:theoryKStokes}~(a) depicts $S_{0}$, which represents the total intensity of the transmitted light. As it is normalised to the output light intensity, its value never exceeds 1. With increasing temperature, the $S_{0}$ absorption profile shows a significant increase in depth and width due to higher atomic density, enhanced Doppler broadening and larger collisional broadening rate. Concurrently, the two $\sigma^{\pm}$ Zeeman absorption dips, at detunings of approximately $\pm2$~GHz, become less resolved, merging into a broader feature centred at 0~GHz due to this increased spectral broadening; this is in agreement with the findings of reference~\cite{Alqarni_2025}.

Figure~\ref{fig:theoryKStokes}~(b) and (c) display $S_{1}$ and $S_{2}$, respectively, which characterise the linear polarisation state of the transmitted light. These parameters describe the rotation of the light's polarisation plane as it propagates through the atomic medium. Their complex, often oscillatory, shapes around resonance (i.e. $\Delta/2\pi = 0~\mathrm{GHz}$) reflect the birefringence (and to a lesser extent dichroism) within the vapour induced by the external magnetic field. The Faraday rotation angle is proportional to the difference in the refractive indices (i.e. the real part of the susceptibility $\chi$) of the right- and left-handed polarisations~\cite{f2f}; as the number density of the vapour grows nearly exponentially with temperature~\cite{pizzey2022laser} there is a rapid evolution of the optical rotation with modest changes in temperature. Notably, at 0 GHz linear detuning (on resonance), $S_{1}$ and $S_{2}$ exhibit distinct values that oscillate between +1 and --1 depending on the vapour cell stem temperature. The rapid frequency dependence of optical rotation, where the absorption spectrum varies much more slowly, is well known for alkali-metal vapours~\cite{siddons2009off, siddons2009gigahertz, kemp2011analytical}. Indeed, the zero-crossings in the $S_{1}$ and $S_{2}$ can be used for laser frequency stabilisation~\cite{mausezahl2024tutorial, krelman2025laser}.

Figure~\ref{fig:theoryKStokes} (d) illustrates $S_{3}$, which quantifies the circular dichrosim of the medium (the difference in absorption of the circular polarisation states) induced by the magnetic field. The extinction of each of the $\sigma^{\pm}$ transitions is described by the corresponding imaginary component of the susceptibility, $\chi$. As the real components of the susceptibility do not play a role in determining either $S_{0}$ or $S_{3}$, we see that these two Stokes parameters do not display the rapid temperature variation that is evident for $S_{1}$ and $S_{2}$. When the Zeeman shift is comparable or larger than the transition linewidth, $S_{3}$ exhibits a characteristic dispersive shape, with a sharp transition through zero at approximately 0 GHz linear detuning. The steepness of this slope provides a measure of the medium's magnetic field sensitivity; as the vapour density increases (i.e. higher temperatures), the gradient of the slope increases.
We note that $S_{3}$ takes values of $\pm 1$ at detunings centred on approximately $\pm2$~GHz. This is because at the detuning of $+(-)$2~GHz the light is resonant with the $\sigma^{+}$ ($\sigma^{-}$) transition; as the medium is optically thick the left-(right-)circular polarisation is completely absorbed, leaving the right-(left)circular component of the incident polarisation state. Consequently $S_{3}$ takes on the value $+1(-1)$. The frequency dependence of the $S_{3}$ signal is used as a discriminant for frequency stabilisation in the dichroic atomic vapour laser lock (DAVLL) technique~\cite{cheron1994laser, corwin1998frequency, overstreet2004zeeman, millett2006davll,reeves2006temperature, nagourney2014quantum}.

Similarly, Figure~\ref{fig:theoryKStokes_magnetic} presents the theoretically predicted Stokes parameters as a function of linear detuning for a 25~mm potassium vapour cell containing 60 Torr of neon buffer gas at fixed temperature of 120$^{\circ}$C, operating under three distinct magnetic fields in Faraday configuration. We notice the same trends as were evident in Figure~\ref{fig:theoryKStokes}: $S_{0}$ and $S_{3}$ exhibit smooth spectral variations, whereas $S_{1}$ and $S_{2}$ -- being dependent on the real part of the susceptibility -- display more complex spectra as a consequence of the optical rotation.

Figures~\ref{fig:theoryKStokes} and \ref{fig:theoryKStokes_magnetic} illustrate the temperature and magnetic field strength dependence of the Stokes parameters in the presence of a known buffer gas filling pressure, from which the shape of the profile can be described. However, the shift and broadening effects, due to the presence of the buffer gas, are not readily apparent. Instead, shown in Figure~\ref{fig:Onresdep}, is a comparison of the four Stokes parameters for vapour cells containing 0~Torr, 60 Torr and 100~Torr of neon buffer gas, with the dependence on vapour cell stem temperature and magnetic field strength plotted.

Shown in Figure~\ref{fig:Onresdep} is the on-resonance (i.e. at zero linear detuning) normalised Stokes parameter values as a function of vapour cell stem temperature for three magnetic field strengths ($B$~=~750~G, 1000~G, and 1200~G). As expected, the $S_{0}$ and $S_{3}$ values evolve smoothly with both temperature and magnetic field, with opposite response (i.e. as the temperature increases $S_{0}\rightarrow0$, whereas $S_{3}\rightarrow1$). Additionally, the vapour cell containing most buffer gas approaches 0 and 1, for $S_{0}$ and $S_{3}$, respectively, at a lower temperature than a vapour cell containing no buffer gas. The presence of the buffer gas induces a line shift; since this $S_{3}$ value is extracted at $\Delta/2\pi$~=~0~GHz, Figure~\ref{fig:Onresdep}~(d) illustrates that $S_{3}$ is a more sensitive method to measure this compared to the other Stokes parameter measurements. 

By contrast there is oscillatory behaviour for both $S_{1}$ and $S_{2}$, with fewer oscillations, within the given temperature range, as the magnetic field strength increases. For vapour cells containing buffer gas, both $S_{1}$ and $S_{2}$ values undergo a damped decay to zero, with a higher buffer gas cell decaying more rapidly for moderate magnetic field strengths (as evidenced by Figure~\ref{fig:Onresdep}~(b)~(i) and (c)~(i)). There is also a difference in temperature from when the medium undergoes these decays to zero. One might assume no optical rotation is occurring when the values of $S_{1}$ and $S_{2}$ are zero, but it is evident from $S_{0}$ -- Figure~\ref{fig:Onresdep}~(a) -- that the medium is optically thick, there is no transmission and this occurs at a lower temperature when there is more buffer gas due to pressure broadening. By increasing the magnetic field, we can reduce ($B$~=~1000~G) or even prevent ($B$~=~1200~G) these decays at the temperature range under investigation. For increasing magnetic field the atomic $\sigma^{\pm}$ resonances are pushed further apart and the refractive index difference (the medium's birefringence) is slightly less on resonance. When the Zeeman splitting becomes dominant over the pressure broadening, we can recover optical transmission and rotation (as evidenced by Figure~\ref{fig:Onresdep}~(b)~(iii) and (c)~(iii) where the three traces are in phase).

Figure~\ref{fig:Onresdep} parts (b) and (c) explain why it is vital to stabilise the temperature of a vapour cell used for off-resonance laser locking, as fluctuations or drifts in temperatures lead to large spectral shifts of the zero-crossing points~\cite{marchant2010off, Zentile_2014}.

Collectively, these plots demonstrate the rich and complex polarimetric signals that arise from the interaction of resonant light with potassium vapour in the presence of a magnetic field and buffer gas, providing a theoretical foundation for interpreting experimental measurements.

\section{Experimental Details}
\label{sec:exp_details}

\begin{figure*}
    \centering
    \includegraphics[width=\linewidth]{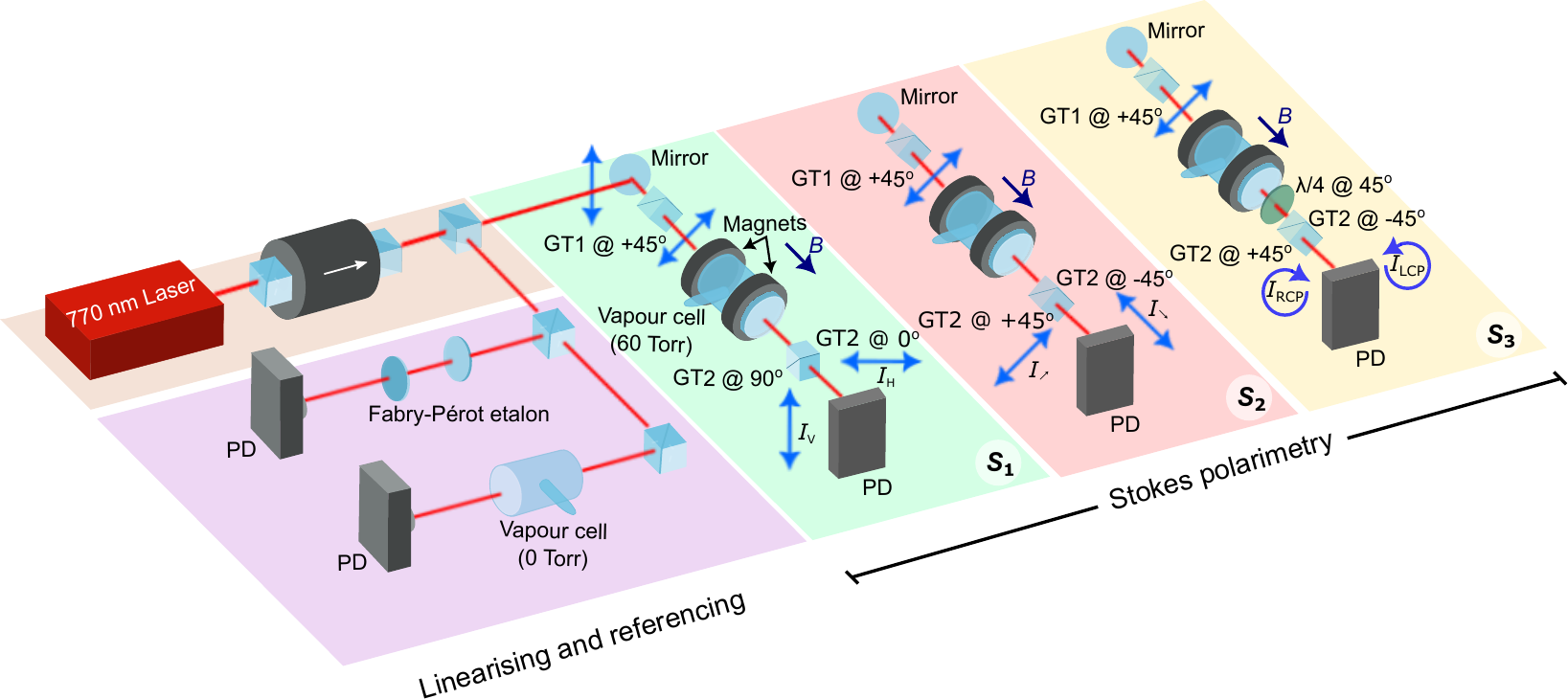}
    \caption{Experimental set-up for potassium Stokes polarimetry. 770 nm laser light is split into two paths: one for linearising and referencing the laser scan (purple box) \cite{pizzey2022laser}, which consists of a Fabry-P\'erot etalon and a potassium vapour cell containing no buffer gas and; a second for performing Stokes polarimetry in a heated 25~mm potassium vapour cell containing 60~Torr of neon buffer gas. The vapour cell is situated in a magnetic field that is in Faraday configuration, which can be tuned between 700--1200~G \cite{pizzey2021tunable}. The experimental set-ups for measuring $S_{1}$, $S_{2}$, and $S_{3}$ are shown in the corresponding boxes labelled ``$S_{1}$", ``$S_{2}$", and ``$S_{3}$", where the first Glan-Taylor polariser (GT1) defines the input polarisation state and the second Glan-Taylor polariser (GT2) measures the output polarisation state on a photodetector (PD). The polarisation state of light is depicted by blue arrows. For each Stokes parameter, two polarisation measurements are taken with the difference resulting in either $S_{1}$, $S_{2}$, or $S_{3}$. $S_{0}$ is determined by summing the two polarisation measurements.}
    \label{fig:setup}    
\end{figure*}

The experimental set-up, shown in Figure~\ref{fig:setup}, utilised a laser beam resonant with the potassium D1 line. The beam was split into two paths: a reference path, used for laser scan linearisation and frequency calibration, and an experimental path, designed to investigate temperature-dependent buffer gas effects and Zeeman splitting on the Stokes parameters.

The reference path remained constant throughout the investigation. It featured a Fabry-Pérot etalon for linearising the laser scan and a 100 mm natural abundance potassium vapour cell -- containing no buffer gas -- that served as a frequency reference \cite{pizzey2022laser}. This reference cell was resistively heated to achieve approximately 50\% absorption, ensuring a stable frequency reference.

The experimental vapour cell was custom-manufactured by Precision Glassblowing, and constructured from 25 mm long fused silica tubes, with flat, parallel, anti-reflection coated windows. To prevent potassium condensation on the windows, a 40 mm stem was incorporated into the cell, which was designed to create a temperature differential. The cell was sealed under 60 Torr of neon gas at room temperature. Two independent resistive cartridge heaters provided precise control over both the stem temperature ($T_{\rm{s}}$) and the cell body temperature ($T_{\rm{c}}$), with both locations monitored by thermocouples. Throughout the experiments, the cell body temperature ($T_{\rm{c}}$) was consistently maintained approximately 20$^{\circ}$C above the stem temperature. An optical power of 800 nW was used for the experimental vapour cell, with corresponding beam waists of \SI[separate-uncertainty]{2.75\pm0.02}{mm}\,×\,\SI[separate-uncertainty]{2.56\pm0.02}{mm}, thereby ensuring operation in the weak-probe regime \cite{weakprobe}. 

A magnetic field was produced using axially magnetised annular permanent magnets, similar to those described in \cite{pizzey2021tunable}, arranged in a Faraday configuration. Here, the magnetic field was parallel to the light's $k$-vector, driving $\sigma^{\pm}$ transitions with left- and right-hand circularly polarised light \cite{f2f}. As the light was not tightly focused, there was no electric field component along the direction of the magnetic field, thus preventing the excitation of $\pi$ transitions \cite{vectorlight}. The annular magnets were designed to accommodate a vapour cell of 25 mm in length (without a stem) and exhibited a root-mean-square (RMS) magnetic field variation of 1--2\% along the cell's length at a maximum magnetic field of 1.4 kG \cite{refId1}. However, with the 6 mm cylindrical stem of our vapour cells, the maximum achievable field was approximately 1.1 kG, with an RMS variation of 3\%. Magnetic field strength was reduced by increasing the separation between the magnets. 

Our investigation covered two primary sets of experiments: (i) measuring Stokes parameters at five different temperatures under a fixed magnetic field; and (ii) examining the impact of three distinct magnetic field strengths (achieved via varying magnet separations) at a constant vapour temperature. The four Stokes parameters were measured using two Glan-Taylor polarisers -- labelled ``GT1'' and ``GT2'' in Figure~\ref{fig:setup} to represent the polariser before and after the vapour cell, respectively -- and a quarter waveplate ($\lambda/4$), which was only inserted when measuring $S_{3}$. 
Linearly polarised light was incident on the first polariser and GT1 was oriented to rotate the plane of polarisation by 45$^{\circ}$, as represented by the blue arrows in Figure~\ref{fig:setup}. Linearly polarised light at this orientation has equal components of both horizontal and vertical linear light, as well as left- and right-hand circular light, making this the ideal polarisation state to measure all four Stokes parameters. 
Due to the way in which Glan-Taylor polarisers work, only the transmitted beam has extremely pure linear polarisation (with the Glan-Taylor having an extinction ratio of 100 000:1). Therefore, to measure the different intensity components for a particular Stokes parameter, the second polariser, GT2, was rotated. For example, to measure the $S_{1}$ parameter, as defined in Eqn.~\ref{eq:Stokesparameters}, GT2 is set to 0$^{\circ}$ when measuring $I_{\text{H}}$ and rotated through 90$^{\circ}$ when measuring $I_{\text{V}}$. Shown in Figure~\ref{fig:setup} are the GT2 angle orientations for each of the Stokes parameter measurements. Before Stokes polarimetry measurements were made, GT1, GT2 and the $\lambda$/4 were calibrated using a simple Malus's law experiment to determine the appropriate angles of orientation. The Stokes parameter results are discussed in Section~\ref{sec:exp_results}.

\section{Experimental Results}
\label{sec:exp_results}

This section presents the experimental results obtained from spectroscopic studies of the potassium D1 line, focussing on two key aspects: the temperature dependence of Stokes parameters in the presence of a buffer gas and the magnetic field dependence. We fitted the experimental spectra using $ElecSus$ by means of a differential evolution algorithm, allowing the stem temperature, cell temperature, shift and broadening to vary. Upper and lower bounds were imposed on $T_{\rm c}$ and $T_{\rm s}$ of $\pm$\,5\,$^{\circ}$C. 

\subsection{\label{subsec:with T}{Temperature dependence of Stokes parameters}}

\begin{figure*}
    \centering
    \includegraphics[width=0.8\linewidth]{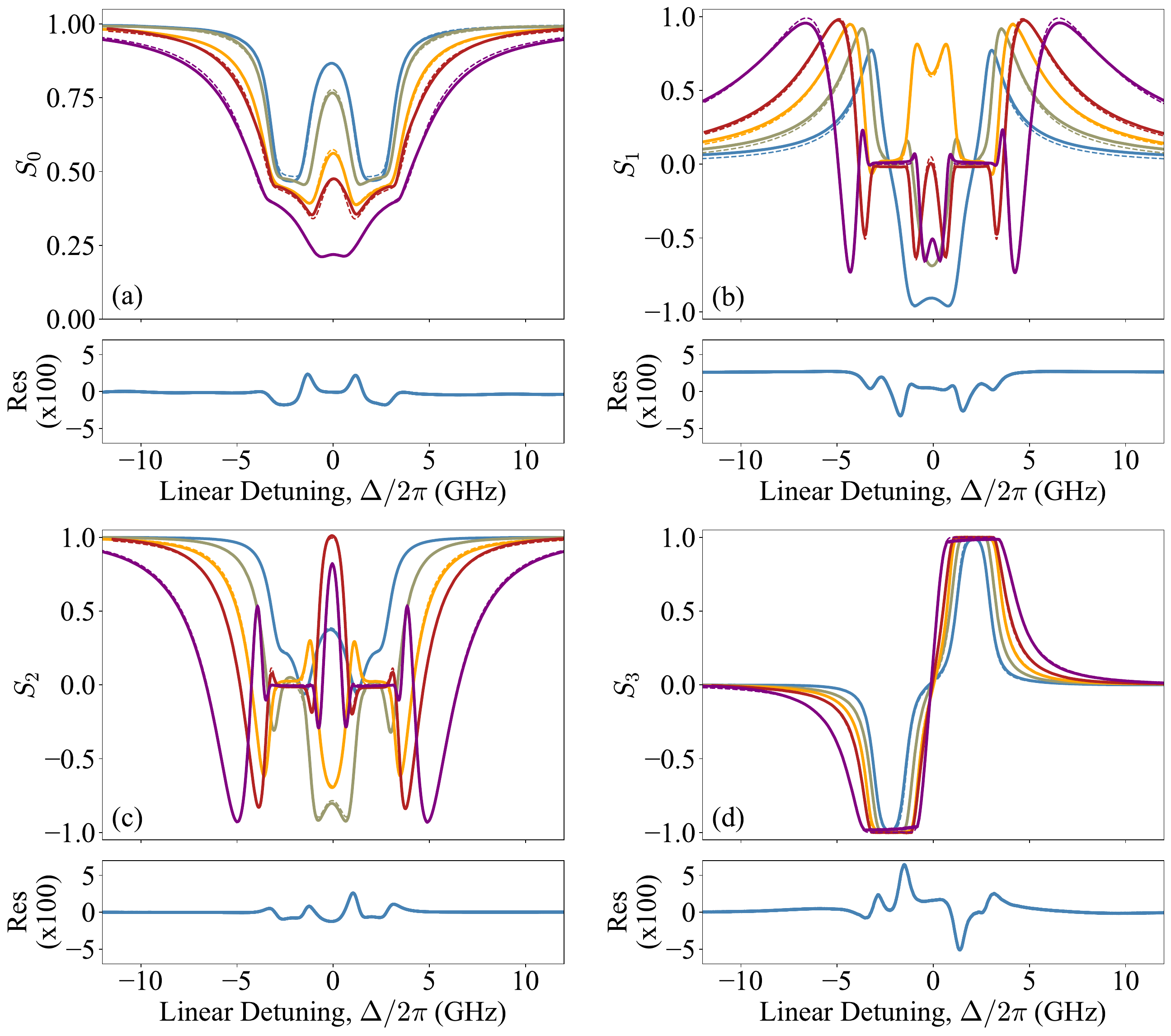}
    \caption{Experimental and theoretical Stokes parameters of the potassium D1 line as a function of linear detuning $\Delta/2\pi$ in a 25~mm natural abundance potassium vapour cell with 60~Torr of neon buffer gas subject to a magnetic field of (1160~$\pm$~4)~G. The solid lines represent the experimental data, while the dashed lines represent the $ElecSus$ theoretical fit to the data. The Stokes parameters shown are (a) $S_{0}$, (b) $S_{1}$, (c) $S_{2}$ and (d) $S_{3}$ for five vapour cell stem temperatures $T_{\rm{s}}$ of 93~$^{\circ}$C (blue), 103~$^{\circ}$C (green), 110~$^{\circ}$C (yellow), 118~$^{\circ}$C (red), and 129~$^{\circ}$C (purple). The residuals, shown below each main subplot, for $T_{\rm{s}}$~=~93~$^{\circ}$C are shown as a demonstration of excellent agreement between experimental data and the $ElecSus$ model.}
    \label{fig:expT}    
\end{figure*}

Figure~\ref{fig:expT} presents the four Stokes parameters as a function of linear detuning measured at a fixed magnetic field strength of (1160~$\pm$~4)~G for a 25~mm vapour cell containing 60~Torr of neon buffer gas. Five temperatures were investigated. Also shown are the $ElecSus$ fits to the data, with the difference in transmission between theory and experiment known as the residuals (multiplied by 100 for better visibility) displayed for one data set underneath. The magnitude and structure of the residuals are used as measures of goodness-of-fit~\cite{hughes2010measurements}.

All of the key spectral features of the Stokes parameters are captured by the theoretical model, which incorporates the effects of the buffer gas presence in the atomic vapour; it is noteworthy that both absorptive features evident in $S_{0}$ and $S_{3}$ and the rapidly varying dispersive features of $S_{1}$ and $S_{2}$ are evidently fully accounted for in the theoretical model.

\subsection{\label{subsec:with B}{Magnetic field dependence of Stokes parameters}}

\begin{figure*}
    \centering
    \includegraphics[width=0.8\linewidth]{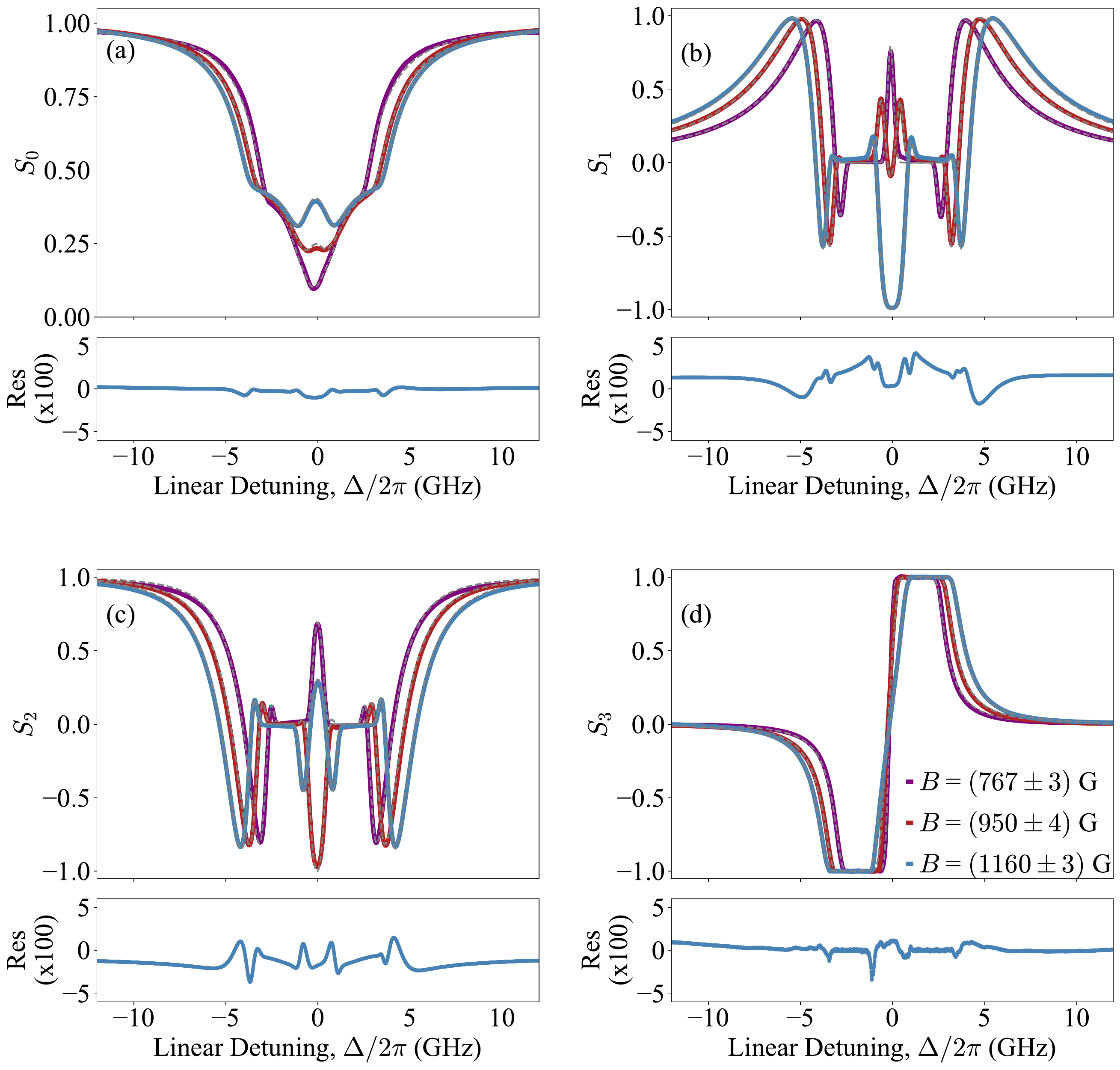}
    \caption{The magnetic field dependence of the four Stokes parameters (a) $S_{0}$, (b) $S_{1}$, (c) $S_{2}$ and (d) $S_{3}$ at a fixed vapour cell stem temperature $T_{\rm{s}}$~=~120~$^{\circ}$C. Shown are the experimental data as well as theoretical $ElecSus$ fitting. Three magnetic fields were investigated: 767~G, 950~G, and 1160~G.}
    \label{fig:expB}    
\end{figure*}

Figure~\ref{fig:expB} illustrates the four Stokes parameters of the potassium D1 line measured at $T_{\rm{s}}$~=~(120~$\pm$~5) $^{\circ}$C for a 25~mm vapour cell containing 60~Torr of neon buffer gas. The spectra are recorded at three different magnetic field strengths that are measured by fitting the spectra to $ElecSus$. The measured magnetic fields are: (767~$\pm$~3)~G; (950~$\pm$~3)~G and (1160~$\pm$~4)~G. Magnetic field measurement errors were determined following the procedure outlined in \cite{Alqarni_2025}: by varying the input polarisation at a fixed magnetic field and fitting $ElecSus$ to extract the magnetic field for each polarisation state. Theoretical modelling of the magnetic field profile at these three permanent magnetic separations gives mean field values in agreement with experiment and predicts axial root-mean-square magnetic field variations of 23\%, 13\% and 3\% over the vapour cell length, respectively. Irrespective of this, the spectra agree well with $ElecSus$, as confirmed by the residuals.

As was observed for the temperature dependence investigation, all of the key spectral features of the Stokes parameters are captured by the theoretical model, which incorporates the effects of the buffer gas presence in the atomic vapour. Once again the rapidly varying dispersive features of $S_{1}$ and $S_{2}$ are evidently fully accounted for in $ElecSus$.

\section{Conclusions}
\label{sec:conclusion}

This study undertook a comprehensive experimental and theoretical investigation into the Stokes polarimetry of potassium (K) atomic vapour confined with neon buffer gas on the D1 line, focusing specifically on the influence of temperature and applied magnetic field. Our primary objective was to characterise how these environmental parameters affect the absorption and polarisation properties of the atomic medium and to validate a theoretical model against experimental observations.

Through measurements in the weak-probe regime, the evolution of the four Stokes parameters ($S_{0}, S_{1}, S_{2}, S_{3}$) was systematically mapped as a function of vapour temperature and magnetic field strength. It was observed that increasing temperature led to a significant increase in absorption depth and spectral broadening for $S_{0}$ with the Zeeman-split components becoming less resolved. Similarly, the linear polarisation signals ($S_{1}$ and $S_{2}$) exhibited increasing complexity, amplitude, and broadening with temperature. The characteristic dispersive profile of $S_{3}$ (circular polarisation), indicative of Faraday rotation, was also shown to vary predictably with both temperature and magnetic field.

A key contribution of this work lies in the successful application and validation of the established $ElecSus$ software package to model these complex phenomena. By incorporating temperature-dependent atomic density, Doppler broadening, and, crucially, the effects of neon buffer gas (pressure broadening and shifts), the theoretical predictions generated by $ElecSus$ showed excellent agreement with our experimental spectra across varying conditions. This work marks the first time $ElecSus$ has been applied to model buffer gas polarimetry of the potassium D1 line, providing valuable benchmark data for future theoretical advancements, and confirming that both the real and imaginary components of the electric susceptibility are fully accounted for in the model.

The findings presented herein significantly enhance our understanding of atom-light interactions in buffer-gas environments. This improved understanding not only refines the predictive capabilities of theoretical frameworks like $ElecSus$ for complex alkali vapour systems but also has implications for diverse applications where such vapours are employed, including precision magnetometry and advanced optical filters.

Future work could extend this investigation to a wider range of neon pressures and temperatures, explore the effects on the potassium D2 line, or investigate the impact of other noble buffer gases to build a more complete picture of collisional interactions. Further refinements to theoretical models to intrinsically account for more complex collisional dynamics could also be pursued, potentially leading to even greater predictive power for advanced quantum technologies.
~\\
~\\
\noindent\textbf{Funding.} IGH acknowledges the funding received from EPSRC (Grant No. EP/R002061/1) and the UK Space Agency (Grant No. UKSAG22\_0031\_ETP2-035). SAA acknowledges support from Najran University, Najran, KSA (Grant No. 443-16-4151). For the purpose of open access, the authors have applied a Creative Commons Attribution (CC BY) licence to any Author Accepted Manuscript version arising from this submission.\\
~\\
~\\
\noindent\textbf{Acknowledgements.} The data that support the findings of this study are openly available at the following URL/DOI: https://doi:10.15128/r1jw827b698.
~\\
~\\
\noindent\textbf{Disclosures.} The authors declare no conflicts of interest.

\bibliographystyle{unsrt}
\bibliography{references}


\end{document}